\documentclass[pra,reprint,singlecolumn,superscriptaddress]{revtex4-1}

\usepackage{tikz}

\usepackage[sort&compress]{natbib}
\usepackage{soul}
\usepackage{graphicx}
\usepackage{epstopdf}
\usepackage{verbatim}
\usepackage{float}
\usepackage{sidecap}
\sidecaptionvpos{figure}{c}
\usepackage{lipsum}
\usepackage{setspace}
\usepackage{dcolumn}
\usepackage{bm}
\usepackage{amssymb}
\usepackage{amsmath}	
\usepackage{mathtools}
\usepackage{color}
\usepackage{multirow}
\usepackage{lipsum}
\usepackage[a4paper]{geometry}
\newgeometry{top=1.5cm,right=1.5cm,left=1.5cm,bottom=1.5cm}
\usepackage[colorlinks=true, urlcolor=purple, pdfborder={0 0 0}]{hyperref}
\hypersetup{
linkcolor=purple,
citecolor=purple
}

\draft

\begin{document}

\title{Measurements and atomistic theory of electron $g$ factor anisotropy\\for phosphorus donors in strained silicon}

\author{M. Usman} \email{musman@unimelb.edu.au} \affiliation{Center for Quantum Computation and Communication Technology, School of Physics, The University of Melbourne, Parkville, 3010, VIC, Australia.}

\author{H. Huebl} \affiliation{Walther-Mei\ss ner-Institut, Bayerische Akademie der Wissenschaften, 85748 Garching, Germany } \affiliation{Physik-Department, Technische Universit\"at M\"unchen, 85748 Garching, Germany} \affiliation{Nanosystems Initiative Munich, 80799 M\"unchen, Germany}

\author{A. R. Stegner} \affiliation{Walter Schottky Institut and Physik-Department, Technische Universit\"at M\"unchen, 85748 Garching, Germany}

\author{C. D. Hill} \affiliation{Center for Quantum Computation and Communication Technology, School of Physics, The University of Melbourne, Parkville, 3010, VIC, Australia.}

\author{M. S. Brandt} \affiliation{Walter Schottky Institut and Physik-Department, Technische Universit\"at M\"unchen, 85748 Garching, Germany}

\author{L. C. L. Hollenberg} \affiliation{Center for Quantum Computation and Communication Technology, School of Physics, The University of Melbourne, Parkville, 3010, VIC, Australia.} 

\begin{abstract}
This work reports the measurement of electron $g$ factor anisotropy ($| \Delta g |$ = $| g_{001} - g_{1 \bar 1 0} |$) for phosphorous donor qubits in strained silicon (sSi = Si/Si$_{1-x}$Ge$_x$) environments. Multi-million-atom tight-binding simulations are performed to understand the measured decrease in $| \Delta g |$ as a function of $x$, which is attributed to a reduction in the interface-related anisotropy. For $x <$7\%, the variation in $| \Delta g |$ is linear and can be described by $\eta_x x$, where $\eta_x \approx$1.62$\times$ 10$^{-3}$. At $x$=20\%, the measured $| \Delta g |$ is 1.2 $\pm$ 0.04 $\times$ 10$^{-3}$, which is in good agreement with the computed value of 1$\times 10^{-3}$. When strain and electric fields are applied simultaneously, the strain effect is predicted to play a dominant role on $| \Delta g |$. Our results provide useful insights on spin properties of sSi:P for spin qubits, and more generally for devices in spintronics and valleytronics areas of research. 
\end{abstract}
\maketitle

\section{Introduction}

Phosphorus impurities in silicon (Si:P) are promising candidates for the implementation of spin-based quantum technologies~\cite{Zwanenburg_RMP_2013, Fuechsle_NN_2012} and quantum computing architectures~\cite{Kane_Nature_1998, Hill_science_2015, Pica_PRB_2016} due to their long coherence times~\cite{Saeedi_Science_2013, Tyryshkin_PRB_2003}. Traditionally the focus has been on electric field control of Si:P nuclear or electron spin qubits~\cite{Kane_Nature_1998}, with remarkable progress towards their fabrication~\cite{Weber_Science_2012} and characterisation~\cite{Pla_Nature_2013, Usman_NN_2016}. Lately the application of mechanical strain has emerged as an alternative control mechanism~\cite{Huebl_PRB_2006, Dreher_PRL_2011, Franke_PRL_2015, Mansir_PRL_2018}. The application of strain is of interest for tuning of the hyperfine interaction~\cite{Huebl_PRB_2006, Dreher_PRL_2011, Mansir_PRL_2018, Usman_PRB_2015} and increase in the exchange interaction coupled with suppression of exchange variations~\cite{Koiller_PRB_2002, Wellard_Hollenberg_PRB_2005}. For control and characterisation of P spin qubits in strained silicon (sSi = Si/Si$_{1-x}$Ge$_x$), one central requirement is to understand the interaction of spins with a strained environment, such as the coupling to orbital degrees of freedom and valley repopulation, which could alter their response to applied magnetic fields. While there has been significant progress on the experimental side in terms of measuring strain-dependent properties of a phosphorus donor atom in sSi~\cite{Wilson_PR_1961, Vrijen_PRA_2000, Huebl_PRB_2006, Dreher_PRL_2011, Franke_PRL_2015, Pla_PRApplied_2018, Mansir_PRL_2018}, the theoretical literature on understanding the spin properties ($g$ factor) of sSi:P is primarily limited to small strain fields ($x \leq$ 1\% or $\varepsilon \leq$ 10$^{-5}$ )~\cite{Roth_PR_1960, Wilson_PR_1961, Mansir_PRL_2018}, whereas a need for larger strain fields (5\% or more) has been predicted to fully exploit the advantage of strain for spin qubit devices~\cite{Wellard_Hollenberg_PRB_2005, Koiller_PRB_2002, Usman_APS_2017}. Furthermore, the existing literature has investigated spin properties of Si:P with the application of electric fields~\cite{Rahman_PRB_2009}, however the electric field dependent variation in $g$ factor for sSi:P system is still an open question. This work reports experimental measurements of the electron $g$ factor in sSi:P samples with strain varying from 7\% to 25\%. Multi-million-atom tight-binding simulations, in good agreement with the measurements, provide key insights in spin properties of sSi:P including the application of electric fields. 

\begin{figure*}
\includegraphics[scale=0.22]{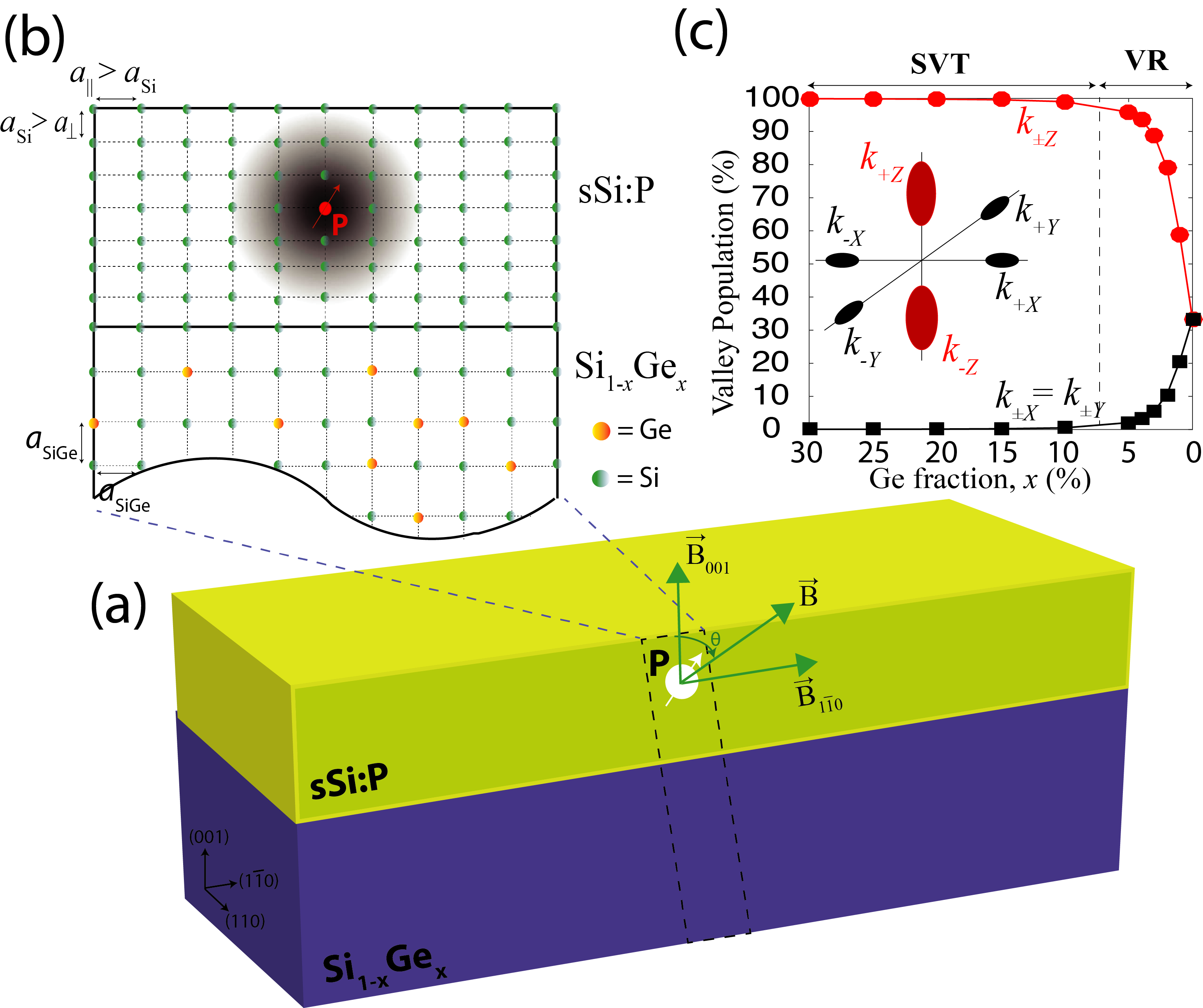}
\caption{\textbf{Schematic diagram of sSi:P qubit device:} (a) The application of strain is implemented by growing a Si:P epi-layer on top of Si$_{1-x}$Ge$_{x}$ substrate, leading to a compressive in-plane strain and a tensile out-of-the-plane strain (see part (b)). The applied magnetic field directions parallel and perpendicular to the strain are also labelled. (c) The valley configuration of the donor ground state A$_1$ as a function of the Ge fraction $x$ in the substrate is plotted, computed from the published analytical model~\cite{Wilson_PR_1961}.}
\label{fig:Fig1}
\end{figure*}

Figure~\ref{fig:Fig1} (a) schematically shows the device structure and the application of strain field is illustrated in (b). The ground state (A$_1$) of unperturbed bulk Si:P is composed of equal contributions from the six degenerate valleys ($\pm k_X, \pm k_Y, \pm k_Z$) at the conduction band minimum of silicon. However, strain breaks the degeneracy of the ground state valley configuration, thereby increasing (decreasing) the population of valley(s) along the compressive (tensile) strain direction. Fig.~\ref{fig:Fig1} (c) plots the valley composition of the donor ground state as a function of strain (given as the Ge fraction $x$ in the substrate) based on simple analytical expressions derived from a valley repopulation model~\cite{Wilson_PR_1961}. Under the application of strain, the population of the $\pm k_Z$ valleys quickly increases and for $x >$ 0.1, the donor ground state is predominantly composed of $\pm k_Z$ valleys. We have labelled $x \leq$ 0.07 as valley repopulation (VR) and $x >$ 0.07 as single-valley-type (SVT) regime of operation. The qubit operation in the SVT regime is important for quantum computing applications as it has been predicted to suppress valley interference-related variations in the exchange interaction~\cite{Koiller_PRB_2002, Wellard_Hollenberg_PRB_2005}. Our results indicate that the $g$ factor anisotropy increases in VR regime and the trend changes in the SVT regime where the anisotropy is found to slightly decrease as a function of strain when the P donor is closer to the interface. When both electric and strain fields are simultaneously applied, the effect of strain plays a dominant role and dictates the strength of $g$ factor anisotropy.           

\section{Experimental measurement of the $g$ factor} 

\begin{figure}
\includegraphics[scale=0.5]{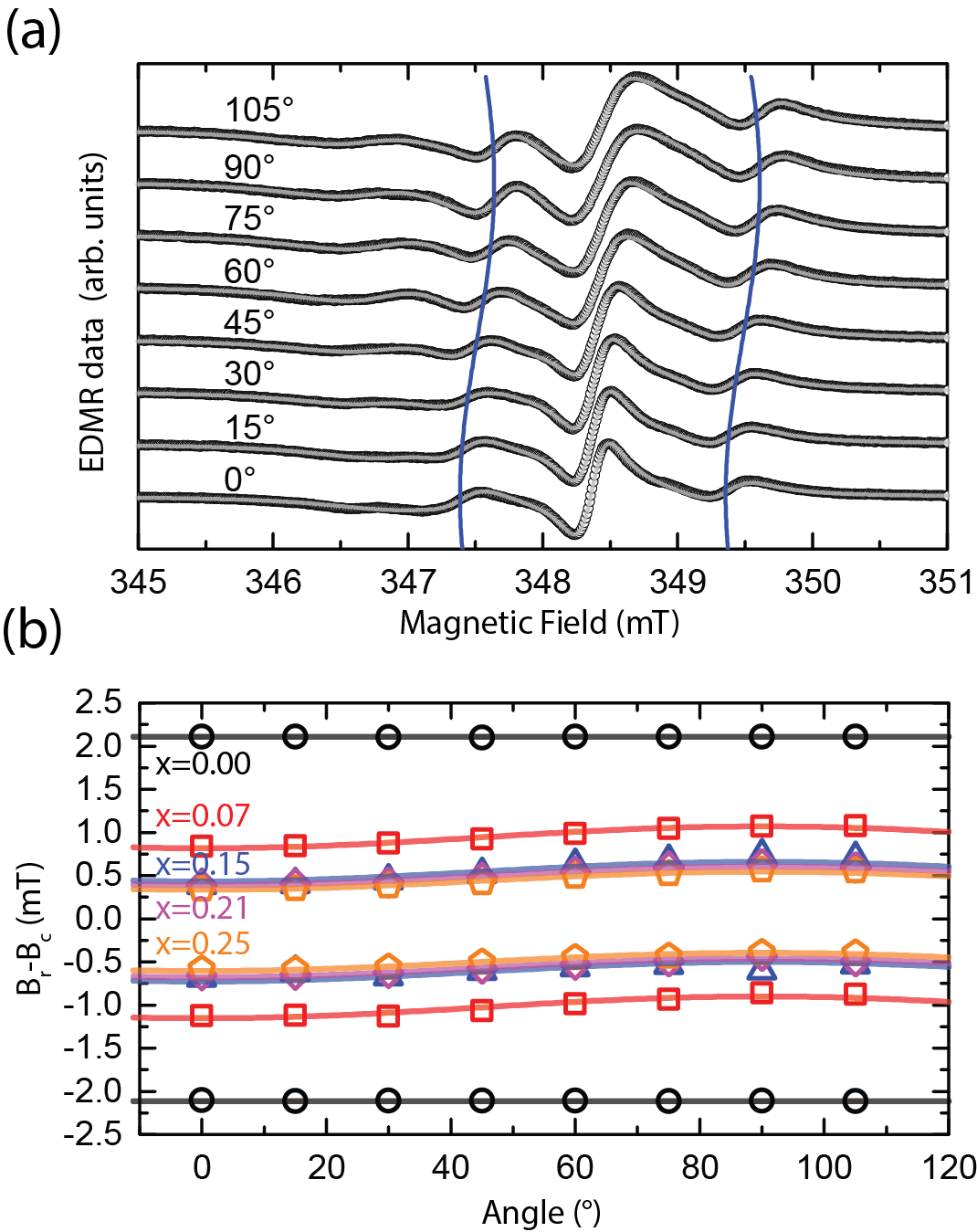}
\caption{\textbf{Experimental measurement of $g$ factor:} (a) Electrically detected magnetic resonance spectra of the phosphorus hyperfine split resonances in compressively stressed silicon grown on a virtual Si$_{0.93}$Ge$_{0.07}$ substrate. (b) Analysed hyperfine splitting and the anisotropy extracted from measurements such as the one shown in (a). For clarity we subtract the center of gravity of the hyperfine splitting B$_c$.}
\label{fig:Fig2}
\end{figure}

Fig.~\ref{fig:Fig2} shows electrically detected magnetic resonance (EDMR) data of fully stressed phosphorus-doped silicon films at $T$ = 5K. The investigated device layer is an in-plane tensile stressed silicon layer with a thickness of 15 nm grown by chemical vapor deposition onto a virtual Si$_{1-x}$Ge$_x$ substrate of 2 $\mu$m thickness. We fabricated strained silicon top layers on various virtual substrates with $x$ up to 0.3. For $x$=0.3, we confirmed the successful tensile stress transfer from the virtual substrate onto the active silicon layer as well as the fully relaxed growth of the Si$_{1-x}$Ge$_x$ virtual substrate using XRD (see Ref.~\cite{Huebl_PRB_2006} for more details). The thin strained silicon layer is doped with phosphorus donors at a concentration of 1$\times$10$^{17}$ cm$^{-3}$. To enable EDMR measurements, we pattern electrical Cr/Au contacts onto the top silicon layer and measure the resistance change of the device under microwave radiation as function of the external magnetic field. To enhance sensitivity, we employ lockin modulation techniques~\cite{Huebl_PRB_2006}.

Figure 2(a) shows spectra for a strained silicon film on Si$_{1-x}$Ge$_x$ with $x$=0.07 (or 7\%) recorded for a rotation of the sample around the (110) axis, where the angle $\theta$ is defined between the (001) axis and the magnetic field direction. We find, besides a reduction of the hyperfine interaction (see Ref.~\cite{Huebl_PRB_2006}), the emergence of a clear anisotropy in the resonance field of the hyperfine lines as indicated by the solid blue lines. In addition to the hyperfine-split peaks originating from the isolated phosphorus donors in the strained silicon host material, we find indications for a central line, which could be attributed to conduction band electrons~\cite{Young_PRB_1997}, as well as a set of lines at lower magnetic fields, which can be identified as the Si/SiO$_2$ interface defect P$_{b0}$~\cite{Poindexter_JAP_1981, Stesmans_JAP_1998}. To obtain information about the $g$ factor anisotropy, we extract the resonance fields $B_r$ of the two hyperfine lines in the spectra for $x$=0, 0.07, 0.15, 0.21, and 0.25 as shown in Fig.~\ref{fig:Fig2} (b), where we have subtracted the field $B_c$ given by the center of gravity of the anisotropy. From this anisotropy data, we obtain the magnitude of anisotropy of the $g$ factor $| \Delta g |$ = $| g_{001} (x) - g_{1 \bar 1 0} (x) |$ as shown in Fig.~\ref{fig:Fig3}(a).

\section{Theoretical calculation of the $g$ factor}

To provide a reliable understanding of the measured $g$ factor, we perform atomistic tight-binding calculations of the P donor wave function with and without the application of strain fields. The Si bulk band structure is reproduced by the $sp^3d^5s^*$ tight-binding model, and the P donor atom is represented by a detailed set of central-cell corrections (CCC)~\cite{Usman_JPCM_2015} and benchmarked against the measured hyperfine values~\cite{Usman_PRB_2015} and high resolution STM images of donor wave function~\cite{Salfi_NatMat_2014, Usman_NN_2016, Usman_Nanoscale_2017}. Based on the tight-binding wave functions of P donor, we then compute the electron $g$ factor by solving the Zeeman Hamiltonian perturbatively~\cite{Rahman_PRB_2009}, where the $g$ factor is computed from the Zeeman splitting of the two lowest spin states. The details of these methods are presented in appendix A. To highlight the dependence of the $g$ factor on strain and the direction of magnetic field $\vec{B}$, we use the notation $g_{\theta} (x)$, where subscript $\theta$ indicates the magnetic field direction and $x$ is the applied strain.

In our model, the application of strain is implemented by increasing the in-plane lattice constants ($a_{100}$ and $a_{010}$) of Si in accordance with the lattice constant of the Si$_{1-x}$Ge$_{x}$ substrate. As a result, the out-of-plane lattice constant ($a_{001}$) undergoes compressive strain in accordance with the Poisson ratio (see Fig.~\ref{fig:Fig1} (b)). Further details of the strain implementation are provided in appendix B. Note that in this paper we define strain in terms of Ge fraction $x$ in the substrate, in contrast to some previous studies where the strain is quantified in terms of valley strain $\chi$ \cite{Wilson_PR_1961, Koiller_PRB_2002} or absolute value $\varepsilon$ \cite{Mansir_PRL_2018}. We should point out that these quantities are directly related to each other and therefore can be used interchangeably. For example, $\chi \sim -$0.98$x$, whereas in-plane $\varepsilon \sim$ 4.2 $\times 10^{-2} x$. For remainder of this paper, we will use $x$ to define the strength of the applied strain field.

\section{Electric field induced anisotropy of the $g$ factor} 

Before computing $g$ factor for the sSi:P case, we first benchmarked our theoretical model against the available electric field-induced Stark shift data of the electron $g$ factor in unstrained Si:P. Table 1 shows the comparison of the calculated and measured $g$ factor Stark shifts, highlighting the excellent agreement of our model with the recent experimental measurements~\cite{Sigillito_PRL_2015} and the theoretical values previously computed from a tight-binding model~\cite{Rahman_PRB_2009}. We note that our theoretical calculations based on the non-static dielectric screening of the donor wave function potential (equation 1 in appendix A) provide a slightly better agreement with the experimentally measured values when compared to the previous tight-binding calculation based on a single value of the silicon dielectric constant.

\begin{table*}[ht!]
\vspace{0ex}
\caption{\label{tab:table2} Computed values of the quadratic Stark shift parameter $\eta_{E}$ for the electron $g$ factor are compared against the experimental and previously reported theoretical values.}
\small{\begin{tabular}
{@{\hspace{0.6ex}}l@{\hspace{0.5cm}}c@{\hspace{0.5cm}}c@{\hspace{0.5cm}}c}
\\
\hline
\hline
     Field & Experiment~\cite{Sigillito_PRL_2015} & Theory (static screening)~\cite{Rahman_PRB_2009} & Theory This Work (non-static screening) \\
		                                Orientations & $\eta_{E}$ ($\mathrm{\mu m^2 / V^2}$)  & $\eta_{E}$ ($\mathrm{\mu m^2 / V^2}$)  & $\eta_{E}$ ($\mathrm{\mu m^2 / V^2}$)  \\
\hline
$\vec{E} \parallel \vec{B}$         & -8 $\pm$ 2 $\times$ 10$^{-6}$ & -12 $\times$ 10$^{-6}$ & -10 $\times$ 10$^{-6}$  \\
$\vec{E} \perp \vec{B}$              & 6 $\pm$ 1.5 $\times$ 10$^{-6}$ & 14 $\times$ 10$^{-6}$ & 8 $\times$ 10$^{-6}$  \\
\hline
\hline
\end{tabular}}
\end{table*}

\section{Strain induced anisotropy of the $g$ factor} 

\begin{figure*}
\includegraphics[scale=0.45]{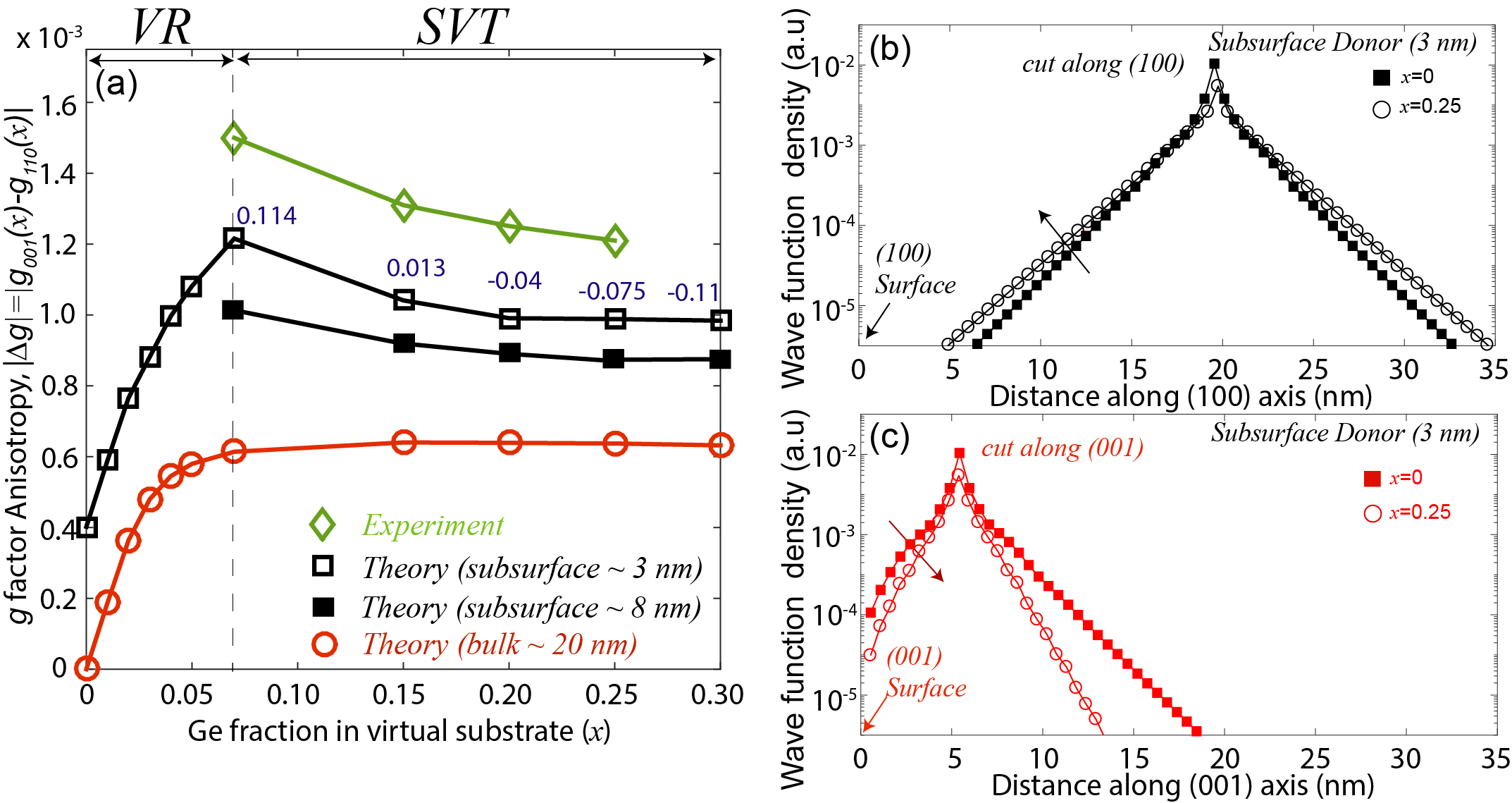}
\caption{\textbf{Strain-induced anisotropy of the $g$ factor:} (a) The measured and theoretically computed $g$ factor anisotropy $| \Delta g |$ is plotted as a function of the substrate Ge fraction $x$. From the simulations, we plot values of $| \Delta g |$ for both bulk and subsurface (3 nm and 8 nm) donor configurations. For 3 nm subsurface configuration, we have also included the values of fractional change ($\Delta$V$^S_{k_Z}$) in the $\pm k_Z$ valley compositions due to interface effect. (b) Line cut of the ground state charge density of P donor wave function for subsurface 3 nm configuration is shown along the (100) axis through the P atom position for the two strain fields corresponding to $x$=0 and $x$=0.25. Only envelope part of the wave function is plotted to indicate its interaction with the interface. (c) Same as (b) but the plot is along the (001) axis.}
\label{fig:Fig3}
\end{figure*}

Fig.~\ref{fig:Fig3} (a) plots the measured and the calculated $g$ factor anisotropies $| \Delta g |$ = $| g_{001} (x) - g_{1 \bar 1 0} (x)|$ as a function of the substrate Ge fraction $x$ in both VR and SVT regimes of strain fields. From theory, we calculate $| \Delta g |$ for three different position configurations of a P donor: a bulk configuration where the distance of P atom from interfaces is larger than 20 nm, and two subsurface configurations where the distance of P atom from (001) interface is 3 nm and 8 nm. In our model, the (001) silicon surface is hydrogen passivated, with the dangling bond energies shifted by a large potential (of the order of 30 eV) to avoid surface states in the energy range of interest~\cite{Lee_PRB_2004}. This creates a large potential barrier at the surface, which blocks any leakage of the wave function outside the boundary of silicon box. 

We first look at the bulk case. In the VR regime, as the applied strain increases, the $\pm k_Z$ ($\pm k_X = \pm k_Y$) valley population of the bulk donor ground state linearly increases (decreases)~\cite{Usman_PRB_2015, Wilson_PR_1961}. This leads to a linear variation in $| \Delta g |$, which was also predicted earlier by effective-mass theory~\cite{Wilson_PR_1961}. This strain dependence of $| \Delta g |$ can be represented by an analytical relation $| \Delta g |$ = $\eta_x x$, where $\eta_x \approx$1.62$\times$ 10$^{-3}$. With the application of the large strain fields in the SVT case, the ground state of the sSi:P donor is predominantly in the $\pm k_Z$ valley state and therefore the $g$ factor converges towards a single valley $g$ factor. We should point out here that a previous calculation of $g$ factor anisotropy based on valley repopulation model has predicted a larger variation ($>$ 10$^{-3}$) for bulk sSi:P~\cite{Wilson_PR_1961}. The valley repopulation model ignores mixing of higher states in the ground state wave function as well as the atomistic representation of the donor wave function and is therefore expected to overestimate the effect of valley reconfiguration. Our tight-binding description takes both of these factors into account and has been shown to exhibit excellent agreement with experimental measurements and DFT calculations of strain dependence of the hyperfine interaction~\cite{Usman_PRB_2015}. 

In the VR regime, the simulated $g$ factor anisotropy for subsurface 3 nm case is shown in Fig.~\ref{fig:Fig3}(a). The interaction of the donor wave function with the (001) surface leads to an asymmetric distribution of the wave function. Furthermore, the ground state has an asymmetric valley compositions ($\pm k_Z > \pm k_X = \pm k_Y$) at $x$=0, which leads to a $g$ factor anisotropy of 0.4$\times$ 10$^{-3}$. The variation of $| \Delta g |$ is linear with $x$, although the slope slightly decrease for strain fields close to the end of VR regime. By fitting of the data in Fig. 3(a), we find that the variation of $| \Delta g |$  can be described by an $\eta_x$ value of $\sim$1.2$\times$ 10$^{-3}$ for small strain fields.      

The experimentally measured data for $g$ factor anisotropy is plotted in Fig.~\ref{fig:Fig3}(a) for $x >$ 0.07. Contrary to the computed bulk $g$ factor anisotropy, the measurements show a small decrease in $| \Delta g |$ when the strain is increased above 7\%. To investigate this effect, we simulate two cases where the P donor is closer to Si interface. These are labelled as subsurface 3 nm and 8 nm in Fig.~\ref{fig:Fig3}(a). In our experimental measurements, the thickness of sSi layer is only 15 nm. Therefore it is expected that the P donor atom should exhibit $| \Delta g |$ variation with strain mediated by significant interface effects. The computed $| \Delta g |$ values for subsurface cases indeed capture the decrease in anisotropy with increasing strain qualitatively following the experimental trend. This decrease of $| \Delta g |$ could be explained by understanding the interplay between the strain and interface effects on the donor ground wave function and its valley composition. The application of strain field for subsurface P donor perturbs the donor ground state in two ways, inducing competing effects on $| \Delta g |$: (i) strain increases $\pm k_Z$ valley compositions and therefore increases $| \Delta g |$, (ii) the compression (elongation) of the spatial distribution of wave function along (001)-axis ((001)-plane) reduces the interface-induced asymmetry of wave function as well as the $\pm k_Z$ valley population. The second effect is shown by plotting line-cuts of donor wave function charge densities in Figs.~\ref{fig:Fig3} (b,c) for the unstrained ($x$=0) and 25\% strain ($x$=0.25) cases, along the two directions: (001) axis and (100) axis thought the donor position. It is clearly evident that the suppression of wave function spatial distribution for $x$=0.25 strain along the (001) axis will reduce the strength of the interface effect.

The two competing effects on donor wave function arising from the interplay between interface and strain produce a net decrease in $\pm k_Z$ valley compositions, in contrast to bulk P donor case where the increase in strain leads to an increase in $\pm k_Z$ valley compositions. To quantitatively provide an estimate of this $\pm k_Z$ valley composition change, we have computed the net change in $\pm k_Z$ valley composition defined by $\Delta$V$^S_{k_Z}$ =  (V$^S_{k_Z} -$ V$^B_{k_Z}$)/V$^B_{k_Z}$, where V$^B_{k_Z}$ and V$^S_{k_Z}$ are $k_Z$ valley compositions for bulk and subsurface 3 nm donor configurations, respectively, at the same applied strain. The values for V$^S_{k_Z}$ are V$^B_{k_Z}$ were computed directly from the donor ground state wave function Fourier spectra in accordance with the published procedure~\cite{Salfi_NatMat_2014}. The values of $\Delta$V$^S_{k_Z}$ are provided in Fig.~\ref{fig:Fig3} (a) for $x >$ 0.07. These values clearly indicate a net decrease in the $\pm k_Z$ valley compositions for subsurface case when the strain is increased. As  $| \Delta g |$ is directly proportional to change in $\Delta$V$^S_{k_Z}$, a decrease in $\Delta$V$_{k_Z}$ is attributed to the observed decrease in $| \Delta g |$ in our measurements.    

\begin{figure*}
\includegraphics[scale=0.35]{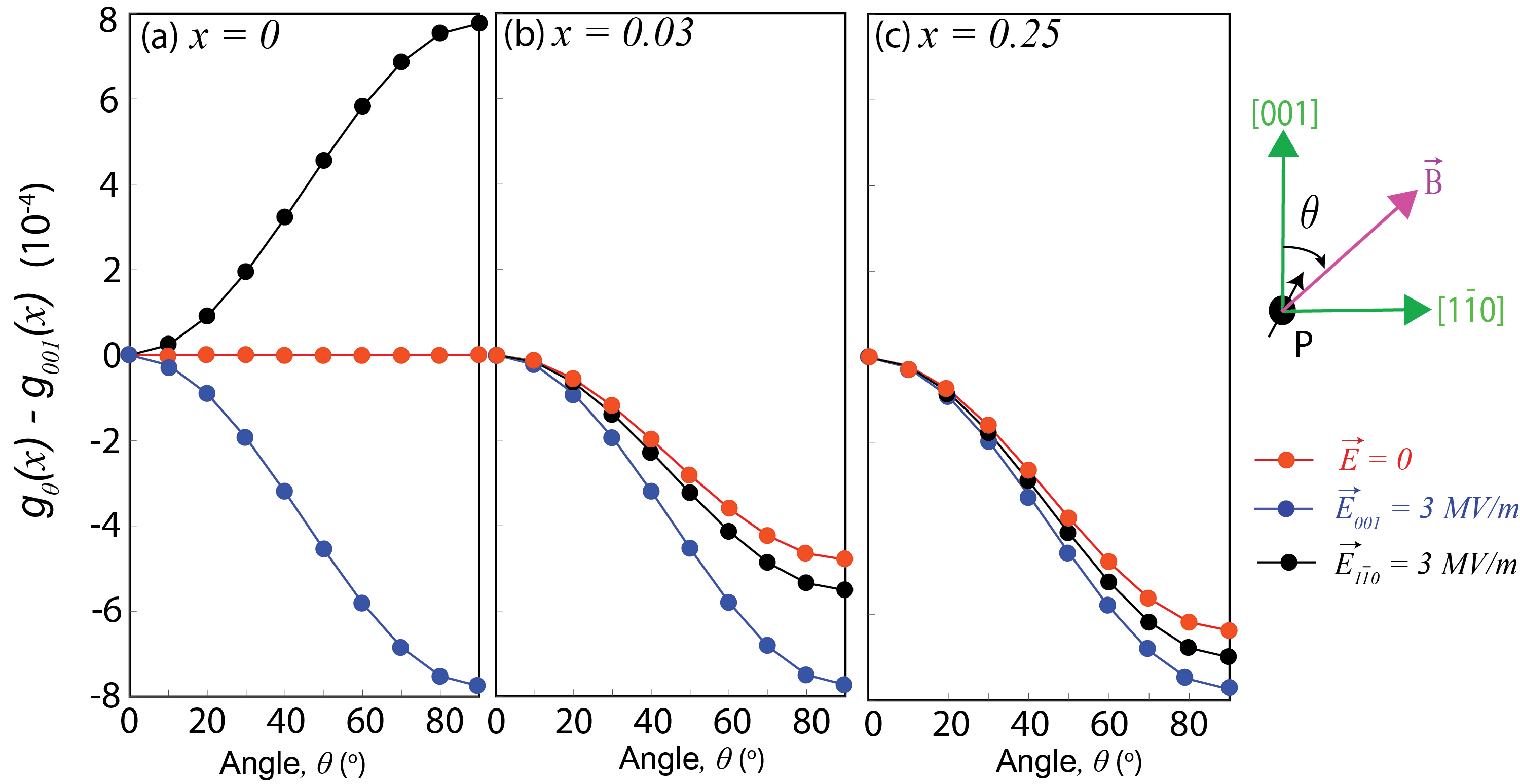}
\caption{\textbf{Rotation of magnetic field:} The $g$ factor anisotropy is plotted as a function of the magnetic field angle $\theta$ with respect to the (001)-axis for three different strain fields ($x$=0,0.03, and 0.25). For each strain field, we investigate three scenarios of the electric field: red circles=no field, blue circles=3 MV/m field along the (001) axis, and black circles=3 MV/m field along the (1$\bar{1}$0) axis.}
\label{fig:Fig4}
\end{figure*}

To summarise our discussion above, the small decrease with strain in $| \Delta g |$ as observed in both experimental measurements and and theoretical calculations for subsurface P donors can be explained as follows: In the VR case, the effect of valley repopulation due to strain is very strong and it overcomes the small decrease in $| \Delta g |$ due to a reduction of the interface effect. In the SVT regime, as the strain effect becomes saturated, the interface effect becomes important and leads to a small reduction in $| \Delta g |$. Although our theoretical results qualitatively follow the measured trend for $| \Delta g |$ dependence on strain in the SVT regime, there is some quantitatively difference as evident from the plots of Fig.~\ref{fig:Fig3}(a). Here we should point out that the experimental measurements were performed for relatively thin (15 nm thick) sSi:P crystal on top of Si$_{1-x}$Ge$_{x}$ substrate. As the P donors are expected to be closer to the sSi/Si$_{1-x}$Ge$_{x}$ interface, significant surface effects are expected in the measured $g$ factor anisotropy~\cite{Vrijen_PRA_2000}. Our simulations do not explicitly include Si$_{1-x}$Ge$_{x}$ substrate, rather only the Si is strained in accordance with $x$. Therefore, we attribute the quantitative discrepancy in $| \Delta g |$ magnitude to the absence of sSi/Si$_{1-x}$Ge$_{x}$ interface in our calculations. Nevertheless the results of our calculations are of the same order of magnitude as measured in the experiment and follow the trend with respect to increasing strain for the subsurface donor case, confirming anisotropy in the magnitude of $g$ factor. Moreover as our simulations accurately includes the net effect of strain on Si:P system, the results will be useful for Si$_{1-x}$Ge$_{x}$ substrate free methods of applying strain to silicon-donor system such as based on silicon-on-insulator (SOI)~\cite{Langdo_APL_2003, Leia_APL_2005} and more recently by using calibrated masses~\cite{Mansir_PRL_2018}.

\section{Effect of magnetic field orientation} 

In Fig.~\ref{fig:Fig4}, we investigate the effect of of the orientation of the magnetic field by varying an angle, $\theta$, from the compressive strain (001) axis to the tensile strain (1$\bar{1}$0) axis. We plot $ \Delta g $=$g_{\theta} (x) - g_{001} (x)$ for a bulk P donor as a function of $\theta$ for three different magnitudes of the applied strain fields: (a) $x$=0, (b) $x$=0.03, and (c) $x$=0.25 corresponding to no strain, VR strain and SVT strain respectively. In each case, we also investigate three scenarios of electric fields as indicated in the figure. For the case of no applied strain (Fig.~\ref{fig:Fig4} (a)), $ \Delta g $ is zero irrespective of the magnetic field orientation when no electric field is applied. This is expected as the $g$ factor is isotropic for bulk P in the absence of any external perturbation. The application of electric field creates a valley repopulation effect by increasing the population of valleys along the electric field axis. For a (001) oriented electric field ($\vec{E}_{001}$), the rotation of $\vec{B}$ field implies a $ \Delta g $ due to $\vec{E} \| \vec{B} - \vec{E} \bot \vec{B}$ case, leading to a negative sign. On the other hand for the (1$\bar{1}$0) electric field, we investigate $\vec{E} \bot \vec{B} - \vec{\textrm E} \| \vec{B}$ case, which is a positive change in $ \Delta g$. These are consistent with the trends observed in table I.

As we turn on a strain field in Figs.~\ref{fig:Fig4} (b) and (c), for $\vec{E}$=0, the anisotropy in $ \Delta g $ increases due to the increase in the valley repopulation as discussed before. Note that $ \Delta g $ will exhibit a linear dependence if plotted against sin$^2 \theta$ (instead of $\theta$) following the relationship $g$=$\sqrt{(g_{||} \textrm cos\theta)^2 + (g_{\perp} \textrm sin\theta)^2}$, which was also previously shown by Wilson et al.~\cite{Wilson_PR_1961}. The application of $\vec{E}_{1 \bar{1} 0}$ has an opposite effect to the strain: the strain shifts higher weight towards $\pm k_Z$ valleys, whereas the in-plane electric field will enhance $\pm k_X$ and $\pm k_Y$ valley populations. From Fig.~\ref{fig:Fig4} (b), we note that even a small strain field (3\%) is sufficient to overcome the electric field effect and reverses the sign of $ \Delta g $. Further increase in the strain to $x$=0.25 adds to the $ \Delta g $ anisotropy.    

The application of an $\vec{E}_{001}$ field increases the valley population of $\pm k_Z$ valleys. Therefore the application of a small strain ($x$=0.03) is sufficient for the donor state to be in the SVT regime. The $ \Delta g $ magnitudes remain same when the strain is increased from 0 to 0.25 in Fig.~\ref{fig:Fig4} (a) to (c). Therefore we conclude that for the sSi:P system, the application of a small strain is sufficient to overcome the effect of in-plane electric fields, whereas SVT behaviour is expected for $(001)$-oriented electric fields even at low strain fields of typical amplitude (3 MV/m).  

\section{A comparison of spin-orbit and hyperfine shifts} 

Recently, it was predicted that a magnetic field of magnitude $\sim$0.78 T makes the Zeeman energy shift due to spin-orbit effects comparable to the hyperfine shift for a bulk unstrained Si:P donors~\cite{Rahman_PRB_2009} under an electric field control. Here we estimate the same quantity for the sSi:P qubits. The ESR frequency shift as described by spin Hamiltonian in a (001) directed magnetic field as: $\Delta \textrm H_z$ = $\Delta g(x) \mu_B \textrm B_z \textrm S_z$ + $\Delta \textrm A(x) \textrm I_z \textrm S_z$, where $\textrm S_z$ and $\textrm I_z$ are the $z$ projections of the electronic and the nuclear spins and $\textrm A(x)$ denotes the hyperfine constant under strain field. For a bulk P donor under large strain ($x$=20\%), the $\Delta \textrm A(x)$ is on the order of 0.25$\textrm A(x=0)$, and $\Delta g(x)$ is on the order of 10$^{-3}g(x=0)$. Using these values, we can estimate the $\textrm B_z$ field on the order of 0.42 T at which the Zeeman shift due to spin-orbit effects becomes comparable to the hyperfine shift under strain control. This is of a similar magnitude as predicted for electric field control and is experimentally realizable. Moreover for $x >$ 15\%, the changes in both $\Delta \textrm A(x)$ and $\Delta g(x)$ are small with respect to further variation in strain, therefore we expect that the requirement for $\textrm B_z$ field will be relatively independent of strain fields in comparison to electric field.  

\section{Conclusions} 

In summary, we have experimentally and theoretically investigated the $g$ factor anisotropy for phosphorus donor qubits in strained Si environments (sSi:P). While the previous theoretical understanding was limited to the application of relatively small strain fields (less than 2\%) restricted to valley repopulation regime of operation, our work probes the $g$ factor anisotropy ($\Delta g$) for both small and large strain fields (varying from 0\% to 30\%) to take advantage of the single-valley-type properties for quantum computing devices. Our results show that for bulk sSi:P system, the linear variation of $\Delta g$ becomes constant at large strain fields. For subsurface donors, the magnitude of the measured $\Delta g$, 1.2 $\pm$ 0.04 $\times$ 10$^{-3}$ is found to be in good agreement with the computed value of 1$\times 10^{-3}$ from multi-million-atom tight-binding simulations explicitly including spin-orbit coupling and central-cell corrections. We also experimentally measure a small decrease in $\Delta g$ magnitude when strain increases above 7\%. We explain this in terms of interface effects which reduce due to deformation of spatial distribution of donor wave function by strain. When electric and strain fields are simultaneously applied, the variation in the $\Delta g$ is dependent on the direction of the electric field with respect to the compressive strain axis. The reported results mark an important step towards understanding magnetic field-dependent spin properties of sSi:P qubits and will be useful for the design and implementation of future quantum technologies.      

\begin{acknowledgements} 
This work is funded by the ARC Center of Excellence for Quantum Computation and Communication Technology (CE1100001027), and in part by the U.S. Army Research Office (W911NF-08-1-0527). HH and MSB acknowledge financial support via the DPG priority programme SPP1601 (HU1896/2 and BR1585/8) and the collaborative research center SFB631. Computational resources from NCN/Nanohub are acknowledged. This work was supported by the computational resources provided by Pawsey Supercomputing Center (Magnus cluster) and National Computational Infrastructure (Raijin cluster) through National Computational Merit Allocation Scheme (NCMAS).
\end{acknowledgements}

\renewcommand{\thefigure}{S\arabic{figure}}
\setcounter{figure}{0}

\setcounter{equation}{0}
\noindent
\\ \\
\textbf{Appendix A. Calculation of donor wave function and the electron $g$ factor}
\\ \\ 
Theoretical investigation of the $g$ factor for sSi:P qubits is limited in the existing literature. In Wilson and Feher~\cite{Wilson_PR_1961}, a valley repopulation model was applied to calculate $g$ factor variation under the application of small strain fields. This simplified model has been shown to exhibit poor agreement for high strain fields with experimental measurement of hyperfine shifts~\cite{Huebl_PRB_2006}, highlighting the need for more sophisticated atomistic approaches such as DFT or tight-binding theory.  To properly understand the anisotropy of the measured $g$ factor, we perform atomistic tight-binding calculations of the donor wave function based on a P atom in a large Si domain (40$\times$40$\times$40 nm$^3$) containing roughly 3.1 million atoms~\cite{Usman_JPCM_2015, Usman_PRB_2015}. The silicon material is represented by a twenty-band \textit{sp$^3$d$^5$s$^*$} tight-binding model, which explicitly incorporates spin-orbit coupling~\cite{Boykin_PRB_2004,Klimeck_2,Ahmed_Enc_2009}. The P donor atom is represented by a Coulomb potential, $U(r)$, which is screened by a non-static dielectric function for Si and is given by:

\begin{equation}
	\label{eq:Nonstatic_donor_potential}
	U \left( r \right) = \frac{-e^2}{ \epsilon  r} \left( 1 + A \epsilon \mathrm{e}^{- \alpha r} + \left( 1-A \right) \epsilon \mathrm{e}^{- \beta r} - \mathrm{e}^{- \gamma r}  \right)
\end{equation}
\normalsize

\noindent
\\ where $e$ is the electronic unit charge and the previously published values of $\epsilon$, $A$, $\alpha$, $\beta$, and $\gamma$ are used~\cite{Usman_JPCM_2015}. The donor potential is truncated to $U_0$ at the donor site, whose value is selected to reproduce the measured $1s$ binding energies. The intrinsic strain in the vicinity of the donor atom is implemented by small nearest-neighbour bond-length deformations predicted by DFT calculations~\cite{Overhof_PRL_2004}. The model has been implemented within the framework of the NEMO3D software package. In the past, this model has demonstrated excellent agreement with the available experimental measurements, such as involving electrical field and strain control of donor hyperfine interactions~\cite{Usman_PRB_2015, Usman_JPCM_2015} and the donor wave function images measured by scanning tunnelling microscope~\cite{Usman_NN_2016, Usman_Nanoscale_2017}. More generally, the tight-binding framework has demonstrated excellent agreement with the experimental measurements on several semiconductor materials and heterostructures~\cite{Usman_1, Neerav_1, Usman_PRB2_2011, Usman_Nanoscale_2015}. 

The calculation of the electron $g$ factor from the donor wave functions is based on solving the Zeeman Hamiltonian perturbatively~\cite{Rahman_PRB_2009} using the matrix elements:

\begin{equation}
	\label{eq:Nonstatic_donor_potential}
	H_{\textrm Zij} = \ < \Psi_i (\vec{r}, x) \mid ( \vec{L} + 2\vec{S}) \cdot \vec{B} \mid \Psi_j (\vec{r},x) >
\end{equation}
\normalsize

\noindent
where $i, j$ represent the spin up/down of the donor states $\Psi$ under the strain field defined by the substrate Ge fraction $x$, and $\vec{L}$ and $\vec{S}$ denote the orbital and spin angular momentum operators, respectively. The $g$ factor is then computed by using the energies $E$ of the two lowest spin states ($\uparrow$ and $\downarrow$) of $H_\textrm Z$: 

\begin{equation}
	\label{eq:Nonstatic_donor_potential}
	g_\theta(x) = \ \frac{(E_{\uparrow} - E_{\downarrow})}{\mu_B | \vec{\textrm B} |}
\end{equation}
\normalsize

\noindent
where $\mu_B$ is the Bohr magneton and $\theta$ is the direction of magnetic field as indicated in Fig.~\ref{fig:Fig1}: $\theta$=0 corresponds to $\vec{B} ||$(001)-axis, $\theta$=90$^o$ to $\vec{B} ||$(1$\bar{1}$0)-axis. 
\noindent
\\ \\
\textbf{Appendix B. Application of strain field}
\\ \\ 
Figure~\ref{fig:Fig1} (a) schematically shows the application of a strain field to a P donor atom in silicon used here~\cite{Huebl_PRB_2006}. The in-plane tensile stressed sSi:P thin film is grown lattice-matched on a virtual Si$_{1-x}$Ge$_{x}$ substrate. The thickness of the sSi:P layer in our samples is chosen to be 15 nm, which is below the critical thickness for strain relaxing defect formation~\cite{Huebl_PRB_2006}. For all Ge fractions $x>$ 0, the lattice constant of Si$_{1-x}$Ge$_{x}$ is greater than the lattice constant of  Si ($a_{\textrm{Si}}$=0.5431 $\mathrm{nm}$). Therefore, the applied stress will stretch the in-plane lattice constant of the Si:P epilayer ($a_{\parallel} > a_{\textrm{Si}}$) and in turn the out-of-plane lattice constant will experience a compressive strain ($a_{\perp} > a_{\textrm{Si}}$) in accordance with the Poisson ratio (see Fig.~\ref{fig:Fig1} (b)). This leads to two inequivalent lattice directions in the strained Si environment: compressive strain along the growth (001) direction and tensile strain in the (001)-plane. 

In some previous theoretical studies~\cite{Wilson_PR_1961, Koiller_PRB_2002}, effective valley strain is used as a parameter to represent the strain effect on donor wave function properties, which is given by:

\small
\begin{equation}
	\label{eq:valley_strain}
	 \chi = \frac{\Xi_u}{3 \Delta_c} \left( \frac{a_{\rm Si} - a_{\rm Ge}}{a_{\rm Si}} \right) \left( 1 + \frac{2 \rm C_{12}}{\rm C_{11}} \right) x  
\end{equation}
\normalsize
\noindent 
Here the value of the uniaxial strain parameter $\Xi_u$ is 8.6 eV, C$_{11}$ and C$_{12}$ are the elastic constants of Si and the value of their ratio C$_{12}$/C$_{11}$ is 2.6, 6$\Delta_c$=12.96 eV is the energy splitting of the singlet (A$_1$) and doublet (E) states for the unstrained bulk P donor, $a_{\rm Si}$=0.5431 $\mathrm{nm}$ and $a_{\rm Ge}$=0.5658 $\mathrm{nm}$ are the bulk Si and Ge lattice constants, respectively, and $x$ is the concentration of Ge in the virtual Si$_{1-x}$Ge$_x$ substrate. This equation shows a direct relationship between $\chi$ and $x$, where $\chi \approx$ $-$0.98$x$, therefore the two representations of strain are interchangeable. In the remainder of this study, we prefer to use $x$ to represent the applied strain, which is a directly relevant experimental parameter.

The strain-induced symmetry breaking has been shown to strongly influence the donor ground state, reducing its binding energy and increasing(decreasing) its $\pm k_Z$ ($\pm k_X, \pm k_Y$) valley contributions~\cite{Usman_PRB_2015}. The ground-state valley configuration is plotted in Fig.~\ref{fig:Fig1} (c) as a function of $x$ in accordance with the published analytical model~\cite{Wilson_PR_1961}. As the strain increases, the $\pm k_Z$-valleys quickly populates and for $x \geq$ 20\%, the ground state is nearly entirely $\pm k_Z$ valley state. We call the $x \leq$ 7\% case as the valley re-population (VR) regime where the population of the $\pm k_Z$-valleys sharply increases with the applied strain. The $x >$ 7\% case is identified as the single-valley-type (SVT) regime because the donor ground state is dominated by $\pm k_Z$-valleys.

In our study, $\vec{B}$ field direction is varied from the (001)-axis towards the (1$\bar 1$0)-axis. Due to the asymmetric population of valleys under a stain field, the impact of the applied magnetic field on the electron $g$ factor is expected to be anisotropic, similar to the previously measured anisotropy in the presence of the applied electric fields~\cite{Sigillito_PRL_2015}. 

\bibliographystyle{apsrev4-1}
%\bibliography{SiDonor}

\end{document}